\documentstyle[prl,aps,psfig]{revtex}
\tighten
\def\be{\begin{equation}}
\def\ee{\end{equation}}
\def\e{\varepsilon}
\def\perpend{\perp}
\begin{document}
\draft
\twocolumn

\title{Localization Transition in Multilayered Disordered Systems}

\author{ S.N. Evangelou$^{1}$, Shi-Jie Xiong$^2$, P. Marko\v s$^3$ and
 D.E. Katsanos$^1$ }

\address{$^1$Department of Physics, University of
Ioannina, Ioannina 451 10, Greece \\
$^2$Department of Physics and National Laboratory of
Solid State Microstructures, \\
Nanjing University, Nanjing 210008, China \\
$^3$ Institute of Physics, Slovak Acad. Sci,
 D\'ubravsk\'a cesta 9, Bratislava 842 28, Slovakia}

\maketitle

\begin{abstract}
The Anderson delocalization-localization transition is studied in
multilayered systems  with randomly placed interlayer bonds
of density $p$ and strength $t$.
In the absence of diagonal disorder ($W=0$),
following an appropriate perturbation expansion, we
estimate the mean free paths in the main directions
and verify by scaling of the conductance that the states remain
extended for any finite $p$, despite the interlayer disorder.
In the presence of additional diagonal disorder ($W > 0$)
we obtain an Anderson transition with critical disorder $W_c$
and localization length exponent $\nu$ independently
of the direction.
The critical conductance distribution $P_{c}(g)$ varies, however,
for the parallel and the perpendicular directions.
The results are discussed in connection to disordered
anisotropic materials.
\end{abstract}
\vspace{1cm}


\section {introduction}

The understanding of the Anderson transition
based on the scaling theory of localization \cite{1}
inspired many detailed  numerical studies
of disordered electronic systems \cite{2}.
The universality of the associated critical
behavior was tested for various  physical models,
which include the crucial role of symmetry \cite {3}
and added  magnetic field \cite{4}.
Universal critical transport properties are also expected
in the presence of hopping matrix elements
which are not the same in the various lattice directions, 
as it can be seen from computations for weakly coupled chains
and coupled planes \cite{5,6}.  It must be emphasized
that many of the previous works on anisotropy include site diagonal
isotropic disorder and involve anisotropy only in the hopping magnitudes. 
This kind of anisotropy remains for zero disorder and is
manifested in the band structure.
However, many realistic materials involve truly anisotropic disorder.
For example, attempting to understand the high-$T_c$ cuprates
within a non-interacting electron picture in the presence
of disorder requires the explanation of the contrasting 
resistivities in the parallel and perpendicular 
directions \cite{7,8,9,10}.  Anisotropic site randomness
in a form resembling a  random superlattice
with lateral inhomogeneities gave anisotropic 
localization for anisotropy below a critical value, 
even for arbitrarily small disorder \cite{11}.

\medskip

The in--plane resistivity for most of the layered high$-T_{c}$
materials exhibits metallic behavior,  increasing linearly with
temperature over a wide temperature range,
while the perpendicular out--of--plane resistivity
is very high at low temperatures and decreases rapidly
as the temperature increases,
reminiscent of semiconductors \cite{12,13,14,15}.
The contrasting behavior of the parallel and the perpendicular
resistivities  was observed in Bi$_2$Sr$_{2-x}$La$_x$CuO$_y$
far below $T_c$, down to the lowest experimental  temperature
\cite{16}. In the underdoped La$_{2-x}$Sr$_x$CuO$_4$
logarithmic divergencies of the corresponding 
resistivities accompanied by a nearly constant anisotropy ratio are,
instead, observed suggesting an unusual three-dimensional ($3D$) insulator.
The electronic transport in these materials 
is expected to arise from scattering in the ``insulating'' layer
between the conducting CuO$_2$ layers\cite{12,13,14,15}.
On the other hand, in almost all high-$T_c$ cuprates
doping impurities or oxygen vacancies
occupy the insulating layers between the
conducting ``pure" CuO$_2$ planes which implies
interlayer disorder.  
Although the main aspects of transport in high-$T_c$ materials,
such as the linear temperature dependence of the in-plane
resistivity, are intimately connected with their strongly correlated nature,
it is believed that anisotropic transport issues are, somehow,
related to their layered structure.

\medskip

We propose a strongly anisotropic multilayered structure 
(see Fig. 1) motivated by realistic anisotropic materials.
This system involves truly anisotropic interlayer disorder,
anisotropic hoppings and the usual isotropic site 
diagonal disorder. It can be regarded as a very simple model 
for the cuprates where the  CuO$_2$ planes 
are believed to be identical without superlattice--like disorder.
Our aim is to study both parallel (${\parallel}$) 
and perpendicular (${\perp}$) transport addressing
the following main questions:
(1) does anisotropic localization occur (for example, 
localization in the layering direction and delocalization
within the layers) for interlayer disorder only?
(2) with additional isotropic diagonal disorder 
is the critical behavior independent of the direction
as scaling theory predicts? Firstly, we compute the conductance 
in the case of interlayer disorder alone
to check whether its localization behavior is the same
in both directions.  Secondly, in the presence of additional 
isotropic disorder of strength $W$ we obtain the critical disorder 
$W_{c}$ and the localization length critical exponent $\nu$ 
to see if they depend on the direction.
We have also analyzed the statistical properties of
the critical conductance  distributions $P_c(g)$  and
although we can conjecture that it is a unique single-parameter function
the distribution for the logarithm of the critical 
$g$ in the ``difficult" ${\perp}$ case
resembles an insulator.

\medskip

In Section II we proceed with the definition of the tight-binding
Hamiltonian for the multilayered lattice structure.
In Section III, we consider the
system with only interlayer disorder and estimate
the corresponding mean free paths in the two directions. 
Our results in the absence of isotropic diagonal disorder 
allow to conclude, in agreement with the scaling
theory, that the states remain extended in both directions 
despite the strongly anisotropic interlayer disorder. However,
the metallic conductance is very different
for the ${\perp}$ case, being insulator-like.
In Section IV we review the numerical methods for
the computation of the conductance in cubic and long wire systems.
We  find singular behavior along the layering direction 
due to the missing bonds. In order to avoid this problem 
we have developed appropriate numerical algorithms 
based on transfer matrix and Green function methods.
Finally, in Section V we discuss the conclusions
of the present study also in connection to realistic
systems.

\section {random multilayer lattice }

We propose a simple $3D$ anisotropic multilayered model 
which consists of parallel lattice planes 
randomly connected by interplane bonds as in Fig. 1
described by the Hamiltonian
\begin{eqnarray}\label{ham}
H &=& \sum_{{\bf m} l} \e_{{\bf m}, l} |{\bf m}, l><{\bf m}, l|
     +\sum_{{\bf \langle m,m^{\prime}\rangle }, l}
     (|{\bf m}, l><{\bf m^{\prime}}, l| \nonumber \\
  & & + t^{\prime}_{{\bf m},l}|{\bf m}, l><{\bf m}, l+1| + \text{H.
c.}),
\end{eqnarray}
where  ${\bf m}$, ${\bf m^{\prime}}$ denote the two-dimensional site
indices in each  layer and $l$ is the layer index.
The first term in Eq. (1) describes diagonal (isotropic)
disorder with the site matrix elements $\e$ chosen randomly
from a box distribution within  $[-W/2,W/2]$,
the second term describes
nearest-neighbor hopping of unit strength within the layers,
which  sets  the energy unit, and
the third term  corresponds to  interplane hoppings
$t^{\prime}_{{\bf m},l}=0$ or $t$,
placed with probability $p$ at random layer positions ${\bf m}$.
The interplane term obeys the binary distribution
\be\label{distribution} P(t^{\prime}_{{\bf m},l})=p\delta
(t^{\prime}_{{\bf m},l},t)+(1-p) \delta (t^{\prime}_{{\bf m},l},0).
\ee
The proposed structure has 
both anisotropic hoppings
due to $t$ and anisotropic disorder due to $p$.  
The transport characteristics are obtained 
by calculating the conductance along the ${\parallel}$ 
and the ${\perp}$ directions.

\medskip

The missing perpendicular bonds in the  layering direction disturb particle
migration even the presence of interlayer disorder alone.
Unlike a naive expectation we find no critical point 
when $W=0$, for any $p \ne 0$. We show that the
system is metallic independently of the direction,
although the behavior of the conductance is very different 
in the two directions.
In the presence of additional diagonal disorder, denoted by $W$,
a critical disorder $W_c$ is obtained for various choices 
of the density $p$ and strength $t$. The critical point $W_c$ 
within finite size errors is found to be the same in both directions.

\section {mean free paths for interlayer disorder }

We consider  anisotropic
disorder in the perpendicular layering direction
represented by $p$ and $t$, due to the
randomly placed bonds among consecutive layers,
in the absence of diagonal disorder  $W$.
In this disordered anisotropic lattice one might expect transport
to be hindered in the perpendicular direction.
It is worth examining whether is present or not.
In order to proceed we adopt a convenient layer-diagonal
representation since for $W=0$ the $2D$ layers are 
perfect planes and can be easily diagonalised.
The eigenstates at the $l-$th layer 
$|{\bf k}_{\parallel}, l\rangle$  are
labelled by the parallel momentum ${\bf k}_{\parallel}$
and the Hamiltonian $H$ can be expressed in the convenient
Bloch-Wannier basis 
\begin{equation}
|{\bf k}_{\parallel}, l\rangle = \frac{1}{\sqrt{N_{\parallel}}}
\sum_{\bf m}e^{i{\bf k}_{\parallel}\cdot {\bf m}}|{\bf m}, l\rangle,
\end{equation}
with parallel momentum ${\bf k}_{\parallel}$, 
the layer index $l$ and ${\bf m}$ summed over 
all $N_{\parallel}=L^2$ sites in every layer for a systen
with $L^3$ sites.  For  $p=0$ the $2D$ layers are perfect
and ${\bf k}_{\parallel}$ is a good quantum number.
For $p=1$ the system reduces to a perfect $3D$
lattice and both ${\bf k}_{\parallel}$,  $k_{\perp}$ 
become good quantum numbers.
 
We consider the case of $p\neq 0, 1$ 
where the translational symmetry in the plane
directions is also broken and ${\bf k}_{\parallel}$ is no longer a
good quantum number. In this mixed representation
the Hamiltonian $H$ can be expressed as
\begin{eqnarray}
H & = & \sum_{{\bf k}_{\parallel},l}\epsilon_{\parallel}({\bf
k}_{\parallel})
|{\bf k}_{\parallel}, l\rangle  \langle {\bf k}_{\parallel},l| \nonumber
\\
 & & + \sum_{l}\sum_{{\bf k}_{\parallel},{\bf k}'_{\parallel}}
[t_{l,l+1}^{{\bf k}_{\parallel},{\bf k}'_{\parallel}}
|{\bf k}_{\parallel}, l\rangle  \langle {\bf k}'_{\parallel},l+1|
+\text{H. c.}],
\end{eqnarray}
with the parallel kinetic energy 
\begin{equation}
\epsilon_{\parallel} ({\bf k}_{\parallel}) =
2\cos (k_{x}) +2\cos(k_{y})
\end{equation}
and the hopping matrix element between neighboring planes
\begin{equation}
t_{l,l+1}^{{\bf k}_{\parallel},{\bf k}'_{\parallel}}
=\frac{1}{N_{\parallel}}\sum_{{\bf m}}e^{i({\bf k}'_{\parallel}-{\bf
k}_{\parallel})\cdot {\bf m}}t'_{l,{\bf m}}.
\end{equation}
In order to investigate the question of localization
in the layering direction we define the retarded
Green function
\begin{equation}
G({\bf k}_{\parallel},l;{\bf k}'_{\parallel},l';t) = -i \theta (t)
\langle [c_{{\bf k}_{\parallel},l}(t),c^{\dag}_{{\bf
k}'_{\parallel},l'}(0)]_+ \rangle ,
\end{equation}
where $c_{{\bf k}_{\parallel},l}(t)$ is the time-dependent destruction
operator of electron in the state $|{\bf k}_{\parallel},l \rangle $.
Its diagonal element $G({\bf k}_{\parallel},l;{\bf k}_{\parallel},l;t)$
gives the probability
for finding an electron on the layer $l$ with  momentum 
${\bf k}_{\parallel}$ at time $t$,
if initially it was on the same layer having the same momentum.
The Fourier transformation of the diagonal Green function 
with respect to time is
\begin{equation}
G({\bf k}_{\parallel},l;{\bf k}_{\parallel},l;E)=
\frac{1}{E-\Sigma ({\bf k}_{\parallel},l,E)}
\end{equation}
with the self-energy $\Sigma_{\parallel}$  written as \cite{17}
\begin{equation}
\label{sig}
\Sigma ({\bf k}_{\parallel},l,E)=
\epsilon_{\parallel} ({\bf k}_{\parallel}) +
\sum_{n=1}^{\infty }\sum_{j}T_j^{(n)}.
\end{equation}
The $n$-th order term is a sum over all paths $j$
of length $n$ starting and
ending in the same state $|{\bf k}_{\parallel},l \rangle $
\begin{equation}
T_j^{(n)}=t_{l,l_1}({\bf k}_{\parallel},{\bf k}_{\parallel 1})
\prod_{i=1}^{n}\frac{t_{l_i,l_{i+1}}^{{\bf k}_{\parallel i},
{\bf k}_{\parallel i+1}}}{E-\epsilon_{\parallel} ({\bf k}_{\parallel
i})}
\end{equation}
with ${\bf k}_{\parallel n+1}={\bf k}_{\parallel}$ and $l_{n+1}=l$.
From Eq. (\ref{distribution})
one has
\begin{equation}
\label{tplane}
t_{l_i,l_{i+1}}^{{\bf k}_{\parallel i},{\bf k}_{\parallel i+1}}
=\frac{t}{N_{\parallel}}\sum_{{\bf m}\in {\cal B}_{l_i,l_{i+1}}}
e^{i({\bf k}_{\parallel i}-{\bf k}_{\parallel i+1})\cdot {\bf m} }
\delta_{l_{i+1},l_i \pm 1},
\end{equation}
with the sum for ${\bf m}$ over the set of sites ${\cal
B}_{l_i,l_{i+1}}$ which  have interplane bond connections.
It is seen from Eq. (\ref{tplane}) that 
the diagonal in momentum matrix elements
(${\bf k}_{\parallel i}= {\bf k}_{\parallel i+1}$)
are  exactly  $pt$. For a given order $n$
if $E$ lies within the pure $2D$ band
$\epsilon_{\parallel}({\bf k}_{\parallel})$
the most divergent term of Eq. (\ref{sig}) comes
from the path which has intermediate states
$|{\bf k}_{\parallel i}, l_i\rangle$
with $\epsilon_{\parallel}({\bf k}_{\parallel i}) \sim E$ for all $i$.
The
corresponding term approaches
$\{ pt/[E-\epsilon_{\parallel}({\bf k}_{\parallel i})]\}^{n}$
which sequentially connects nearest neighbor plane states with the
same ${\bf k}_{\parallel i}$.
It can be also seen that momentum scattering is always accompanied by
interlayer hopping since  a change of ${\bf k}_{\parallel}$
leads to a change of the layer index.

\medskip

The above analysis splits the Hamiltonian 
into two parts $H=H_0 + H_1$.  The ``undisturbed" part $H_0$
represents a perfect anisotropic $3D$ lattice
with intralayer (interlayer) hopping $1$ ($pt$).
The rest is a ``random'' Hamiltonian $H_1$
with only off-diagonal matrix elements
expressed in the $H_0$-diagonal basis 
$|{\bf k}\rangle = |{\bf k}_{\parallel},k_{z}\rangle$ of the form
\begin{eqnarray}
\label{h1ele}
H_1({\bf k},{\bf k}') & = &
\frac{t(e^{ik_z}+e^{-ik'_z})}{N}\sum_{l,{\bf m}\in {\cal B}_{l,l+1}}
e^{i({\bf k}_{\parallel}-{\bf k}_{\parallel}')\cdot {\bf m}
+i(k_z-k_z')l},
\nonumber \\
 & & \text{ for } {\bf k}_{\parallel }\neq {\bf k}_{\parallel }',
\nonumber \\
H_1({\bf k},{\bf k}') & = & 0, \text{ for } {\bf k}_{\parallel } = {\bf
k}_{\parallel }',
\end{eqnarray}
where $N$ is the total number of lattice sites and
$k_z$ the perpendicular momentum.

\medskip

It can be seen from Eq. (\ref{h1ele}) that the matrix elements of
$H_1$ are complex numbers of average amplitude value modulo 1
plus a  random phase. 
If the size of the system increases to infinity the phase exhausts all
possible values in $[0,2\pi ]$ and the average should vanish. In this
situation the scattering by the random configurations of the interplane
bonds can be well described by perturbation theory with 
${\bf k}$-space self-energy 
\begin{equation}
\Sigma ({\bf k},E) \simeq \epsilon ({\bf k}) +
{\sum_{\bf k'}} \frac {|H_1({\bf k, k'})|^{2}}
{E-\epsilon ({\bf k}')-i0^+}  .
\end{equation}
The configuration average
$ \langle \text{Im} {\sum_{\bf k'}} {\frac  {|H_1({\bf k, k'})|^{2}}
{E-\epsilon ({\bf k}') -i \Gamma }} \rangle$
can be computed as a function of $p$ and the 
results fitted to a semicircular form as
$\rho(E) t^{2} p(1-p)$, with $E$ within the
$H_0$ band and $\rho (E)$ the corresponding 
density of states.  This allows to estimate the
lifetime of states  $\tau \sim \frac{1}{\rho (E)t^{2}p(1-p)}$
and the corresponding mean free paths
\begin{equation}
\lambda_{\parallel} = {\frac {\tau u_{\parallel}^{2}}{u}} \simeq
\frac{1} {\rho (E)t^{2}p(1-p)\sqrt{2+p^{2}t^{2}}},
\end{equation}
\begin{equation}
\lambda_{\perp} = {\frac {\tau u_{\perp}^{2}}{u}} \simeq
\frac{p} {\rho (E)(1-p)\sqrt{2+p^{2}t^{2}}},
\end{equation}
for Fermi velocites $u_{\parallel (\perp)}$.

\medskip

We observe that for small-$p$ the obtained mean free path 
in the parallel (perpendicular) direction
is proportional to $1/p$ ($p$).
This implies that the scattering strength of the interlayer 
disorder increases with $p$ for parallel transport
but decreases with $p$ for perpendicular transport. 
This intrinsically anisotropic situation seems in contradiction 
with the scaling theory of localization
because $p\rightarrow 0$ might be thought to be in favor of localization 
in the perpendicular direction, while transport 
is the least affected giving extended states  
in the parallel direction.
However, the application of the one-parameter scaling
theory to this situation should give a common critical point in all directions 
so that transport in the perpendicular direction should be extended
as well. This is, indeed, numerically confirmed
in Chapter IV where we show  
that the scaling behavior of the 
conductance in finite cubic systems with interlayer
disorder is in agreement with the scaling theory.
It turns out that the interlayer disorder is not sufficient
to localize the electrons, even in the ``difficult" layering direction,
without any additional diagonal disorder $W$. 
The statistical behavior of the conductance, however, 
is very different in the two directions.  

\section {numerical calculation of the conductance}

\subsection{The cube }

The parallel (perpendicular) conductance  
${\frac {e^2}{h}} g_{\parallel (\perp )}(L)$
can be obtained for a  $L\times L\times L$ cubic system 
at the Fermi energy $E$ directly from the multichannel 
Landauer--Buttiker formula \cite{18}
\begin{equation}
g_{\parallel (\perp )}(L) = \text{Tr}({\bf t}^+_{\parallel
(\perp )}{\bf t}_{\parallel (\perp )}), 
\end{equation}
where ${\bf t}_{\parallel (\perp )}$ is the 
transmission matrix for electronic
propagation along the ${\parallel}$ (${\perp}$) direction
computed by transfer matrix techniques. Two perfect
semi-infinite bars are attached to two opposite sides of this cube 
and hard-wall boundary conditions are used for the other sides. 
The number of independent channels for the incoming and outgoing
leads is $L^2$ and the transmission matrix can be calculated from
the amplitudes of transmitted waves in the outgoing leads
by assigning, at a time, a unit incident wave amplitude
for one channel in the incoming leads and zero for the rest.
We can easily establish the
recursion relations for the corresponding
wave function coefficients along the parallel direction
to calculate $g_{\parallel}$.
In the ${\perp}$ direction
computing the matrix ${\bf t}_{\perp}$ is not possible
since the recursion relations are singular
due to the presence of zero hoppings
for the missing interplane bonds.
To overcome this difficulty we set up the recursion relations 
along the direction parallel to the planes, perpendicular to the
leads, but with the reflection and transmission
coefficients in the channels as the unknown variables. 
We can solve these recursion relations with hard wall
boundary conditions perpendicular 
to the leads. By this method we  obtain 
the perpendicular transmission matrix ${\bf t}_{\perp}$
avoiding the singularities due to missing bonds.
This is a convenient tool to consider
propagation in the layering $\perp$ direction by a transfer matrix
product only along the easy ${\parallel}$ direction.
In order to suppress fluctuations
we have taken averages over up to a $5000$ random 
cubic configurations in each case \cite {19}.

\subsection{The wire}

We can also compute the parallel (perpendicular) dimensionless
conductance  $g_{\parallel (\perp )}(M)$
for a quasi-one-dimensional $ M\times M\times L$ geometry,
via Green function methods \cite{20}.
In the parallel direction ${\bf m}=x,y$
the Hamiltonian of the $M\times M$ slice
is incorporated into the transfer matrix
\be\label{transfer}
T_x=
\left(\begin{array}{cc}
V_{x,x+1}^{-1} & 0\\
0         & 1
\end{array}
\right)
\left(\begin{array}{cc}
E-H_x & -1\\
1 & 0
\end{array}
\right)
\left(\begin{array}{cc}
1 & 0\\
0 & V_{x-1,x}
\end{array}
\right),
\ee
where the matrix $V_{x,x+1}$ has 
unit elements.  For large length  $L$ in the $x$-direction the product
\be
T=\prod_{x=1}^{L} T_x,
\ee
has eigenvalues $\exp \gamma_i L$
and Lyapunov exponents $\gamma_i$,
$i=1,2,\dots,M^2$. The smallest positive
Lyapunov exponent $\gamma_1$ determines the
scaling parameter $\Lambda_\parallel$, via
\be
\Lambda^{-1}_\parallel=M\gamma_1={\left(\frac{\xi_M}{M}\right)}^{-1},
\ee
with $\xi_M=\frac{1}{\gamma_1}$ the largest localization length,
which is of interest for finite-size scaling studies when
the width $M^2$ of the slice is varied.
We obtain the critical value $\Lambda_{c\parallel}$ at the point where 
$\Lambda_{\parallel}$ becomes independent of $M$ (see Table 1).
In order to ensure accuracy of about 1\% for $\Lambda_\parallel$ the 
length of the studied system is more than about $200000$.
We find that this length near the critical point varies
as $\propto \Lambda^{-1}$. 
In the case of very strong anisotropy we were unable to  obtain this 
accuracy for all $M$ and $W$.

\medskip

This method cannot be used  in the $\perp$-direction. The reason is,
again, the  zero elements of the hopping matrix $V_{l,l+1}$
with probability $1-p$ so that the
inverse matrix $V_{l,l+1}^{-1}$ which enters (\ref{transfer})
becomes singular. In order to avoid this problem
we can alternatively use the Green function
$G(E)=(E-H)^{-1}$ by applying the iterative scheme of \cite{20}
based on the two equations
\be\label{gf1}
G_{1,l+1}^{(l+1)}=G_{1,l}^{(l)}V_{l,l+1}G_{l+1,l+1}^{(l+1)}
\ee
and
\be\label{gf2}
G_{l+1,l+1}^{(l+1)}=[E-H_{l+1}-V_{l+1,l}G_{l,l}^{(l)}V_{l,l+1}]^{-1}.
\ee
The Hamiltonian  $H_l$ represents the $l$th  $M\times M$ layer,
$V_{l,l+1}$ the hopping between layers $l$, $l+1$ and
$G_{l_1,l_2}^{(l)} $ is the Green function of the system with length $l$
between layers $l_1$ and $l_2$. The diagonal matrix $V_{l,l+1}$
of the order of $M^2 $ has zeros  with probability $1-p$
so that $V_{l,l+1}^{-1}$ becomes singular.
The main advantage of the formulae (\ref{gf1},\ref{gf2})
is that they do not contain the inverse of  $V_{l,l+1}$.
The disadvantage is the necessity to invert a matrix
of order $M^{2}$ [Eq. (\ref{gf2})] at each iteration step.
In this case we restrict the number of iterations to $L\le 40000$
and the corresponding peprendicular scaling
parameter is defined as
\be
$$\Lambda_{\perp}^{-1}=\frac{M}{L}\log {\rm Tr} G_{1L}^L=
{\left(\frac{\xi_M}{M}\right)}^{-1}.$$
\ee

\medskip

The critical disorder $W_c$ and the critical exponent $\nu$ are computed 
in both directions from the numerical data of $\Lambda(M,W)$.
The usual linearization near the critical point
\be\label{Lambda}
\log\Lambda(M,W)=\alpha_M +\beta_M\log W,
\ee
and the independence of $\Lambda$ on $M$ 
at the critical point gives $W_c$ from the slope of
the linear dependence
\be\label{alpha}
\alpha_M=-\log W_c\times \beta_M +{\rm const}.
\ee
The obtained critical $\Lambda_{c\parallel}$
and $\Lambda_{c\perp}$ are different
from the value $\Lambda_c$ of the corresponding isotropic system
(see however Eq. (\ref{lambdaiso}) below).  
The critical exponent $\nu$ is
determined from the coefficients $\beta_{M}$ via
\be\label{nu}
\nu=-{\log M\over \log\beta_M}.
\ee
We have also repeated the computations for 
cubic systems and although the finite size effects
became more pronounced the obtained critical values
differ very little from those obtained for long wires.

\medskip

In agreement with the $W=0$ case
it is reasonable to suppose  that the form of $P_c(g)$
also depends  on the direction.
For the $\parallel$ direction the conductance is calculated from
the formula
\be\label{conductance}
g(M)=\sum_{i=1}^{M^2}\frac{2}{\cosh^2(z_i/2)}
\ee
where the $z_i$'s are the logarithms of the $i$th eigenvalues of
the matrix ${\bf t}^{\dag}{\bf t}$ \cite{21} and
in the limit $L>>M$ converge to $2M \gamma_i$.
In the $\perp$ direction the
singular behavior of the matrix $V_{l,l+1}$ does not permit to use
the formula (\ref{conductance}).
To calculate $\log g_{\perp}$ in this case we use the fact that
$\Lambda_{c\perp}<<1$ for most of the critical points
discussed below.  Then, the critical conductance from 
$z_1=2/\Lambda_c$ is very small
and  can be estimated from the contribution of the first channel via
\be\label{gtwoo}
\log g_\perp = {\rm Tr}~ G_{1,L}(E+i\eta)G_{L,1}(E-i\eta),
\ee
with the imaginary part of the energy $\eta$ given by the
ratio of the bandwidth over the mean level spacing \cite{22}.

\section{results}

\subsection {$ W=0 $}

Fig. 2 shows the scaling behavior  of
$g_{\parallel (\perp )}(L)$ for the $L^3$-site cubic system
for various bond densities $p$ and
anisotropic interplane coupling $t=0.3$,
in the absence of diagonal disorder $W$.
The parallel conductance is shown to increase ballistically
($\sim L^{2}$) for small $L$ and linearly for higher $L$.
In the large size ($L\to \infty$) limit  the
corresponding scaling function $\beta (L) = d\ln g/d\ln L$
becomes positive for $g_{\parallel (\perp)}$,
which implies extended states in both  directions
for any  $p$, in agreement  with the  scaling theory
which predicts a common critical point in any direction.
However, the obtained transport behavior is essentially different
in the two directions.
In Fig. 3 we show the energy dependence of the conductance
for  $p=0.5$ and $t=0.3$ where
the ratio of the two conductances is close to the estimate
$g_{\parallel }/g_{\perp }\approx (t_{\parallel}/t_{\perp})^2$
for $t_{\parallel}=1$ and $t_{\perp}=pt=0.15$ \cite{6}.
A key  finding  from Fig. 2 is a rather smooth
$g_{\parallel}(E)$  while $g_{\perp}(E)$ displays violent
oscillations as a function of $E$.
The dips in $g_{\perp}(E)$ can be regarded as due to ``minigaps"
in the perpendicular direction
which might have effect similar to a semiconductor,
leading to  insulating kind of behavior for the
out-of-plane conductivity when the Fermi energy 
is varied.

\subsection{$ W>0 $}

The $W$-dependence of $\Lambda(M,W)$ for different $M$'s
and various parameters $p,t$ is presented in Figs 4., 5., 6., 7. 
The  corresponding critical points are calculated
by the described fitting procedure.
The  results are listed in Table 1 and
the data are very reliable for the $\parallel$ direction.
In the $\perp$ direction they are much harder to analyze due
to finite-size effects. For example, in case
D (Fig. 7) the obtained $M$-dependence of $\Lambda_\perp(M)$
is not monotonic and for $W=5$ we find  $\Lambda(M)$
which decreases with $M$ for small system sizes,
imitating insulating behavior. However, for larger
$M>12$ the values of $\Lambda (M)$ begin to increase and the correct 
scaling is restored.
This  $M$-dependence  is caused by a second
irrelevant scaling parameter from the relation
$\Lambda= aM^{1/\nu}+bM^\beta$ with $\beta<0$.
The correct estimation of the critical parameters in the
${\perp}$ direction requires either
numerical data for larger $M$ or  possibly
more sophisticated fits \cite{23}.
Although for strong anisotropy 
we could not  obtain accurately the critical parameters in the
$\perp$ direction we check  
that our results converge to those of the
$\parallel$ direction when $M$ grows.
\begin{table}[t]
\begin{center}
\begin{tabular}{|c|c|c|c|c|c|c|c|c|}
\hline
  & $p$ & $t$ & $W_c$ & $\Lambda_c$ & $\nu$ & $M$ &
$\langle g\rangle$ & \multicolumn{1}{c|}{var $g$}   \\
\hline
%
 \multicolumn{9}{c}{ $\parallel$ }  \\
\hline
A & 0.6 & 1.0&  14.47 $\pm$ 1.10 & 0.676 & 1.619 $\pm$ 0.14 & 10-16 & 0.59 &
 \multicolumn{1}{c|}{0.16}  \\
\cline{1-8} \cline{8-9}
B & 0.6 & 0.3 & 10.48 $\pm$ 0.58 & 0.933 & 1.666 $\pm$ 0.10 & 6-14  & 1.5  &
\multicolumn{1}{c|}{0.4}    \\
\hline
C & 0.1 & 0.3  & 7.93 $\pm$ 0.24 & 1.218 & 1.514 $\pm$ 0.06 & 12-18 & 2.6  &
 \multicolumn{1}{c|}{0.7}    \\
\hline
D & 0.6 & 0.1  & 8.05 $\pm$ 0.45 & 1.174 & 1.790 $\pm$ 0.14 & 6-16  & 2.6  &
 \multicolumn{1}{c|}{0.7}    \\
\hline
%
 \multicolumn{9}{c}{ $\perpend$ }  \\
\hline
A & 0.6 & 1.0 &14.30 $\pm$ 1.30 & 0.437 & 1.546 $\pm$ 0.16 & 6-14  & &
 \multicolumn{1}{c|}{} \\
\hline
B & 0.6 & 0.3 &10.20 $\pm$ 0.67 & 0.189 & 1.308 $\pm$ 0.21 & 8-18  & &
 \multicolumn{1}{c|}{} \\
\hline
C & 0.1 & 0.3 & 7.18 $\pm$ 0.74 & 0.089 &        -         & 12-20 & &
 \multicolumn{1}{c|}{} \\
\hline
D & 0.6 & 0.1  &6.80 $\pm 1.94$ & 0.100 &        -         & 14-18 & &
 \multicolumn{1}{c|}{} \\
\hline
\end{tabular}
\end{center}
\caption{Review of various critical parameters with 
the mean conductance $\langle g\rangle$ and its variance
var $g=\langle g^2\rangle -\langle g\rangle^2$.
The obtained values confirm that the critical disorder
is very close for ${\parallel}$ and ${\perp}$ transport,
in agreement with the one-parameter scaling theory.
We could not calculate the critical exponent $\nu$ 
in  the ${\perp}$ direction for the cases C, D.}
\end{table}
\noindent
In this direction the critical region is very narrow and we
could neither calculate the critical exponent,
since  larger system sizes are required.
Nevertheless, the scaling analysis for
A and B gives satisfactory results in both
directions which confirm the equal $W_c$ and $\nu$. 
Moreover, we find the quantity
\be\label{lambdaiso}
\Lambda_{c}=\big[\Lambda^2_{c\parallel}
\times\Lambda_{c\perpend}\big]^{1/3}
\ee
which gives the critical value of the corresponding isotropic model
\cite{6}.

Fig. 8(a), (b) presents the probability distribution
of the critical conductance $P_c(g_{\parallel})$
in the parallel direction. 
The distribution is shown to be size-independent but
depends on the various critical points.
For C and D the obtained $P_c(g_{\parallel})$
and $\Lambda_{c\parallel}$ are the same. This indicates
that  $P_c(g_{\parallel})$ is determined only 
by $\Lambda_{c\parallel}$ \cite{21}.
The numerical data for the mean and variance of the conductance 
are also unique functions of $\Lambda_{c\parallel}$ 
(see Table 1) supporting this conjecture.  
In the limit $t\to 0$ the critical
distribution in the parallel direction is expected
to converge to a Gaussian. However, the 
spectrum of the obtained higher Lyapunov exponents shows square-root behavior 
similar to isotropic $3D$ disordered systems \cite{22}.
To display the dramatic differences in parallel
and perpendicular transport, 
we also present the critical distribution of $\log g_{\perp}$, 
calculated for the perpendicular direction for the critical
case C. This distribution has all the features of the 
localized regime since it is log-normal with 
\be\label{var}
{\rm var} \log g_\perp \approx -\langle\log g_\perp\rangle.
\ee
The important difference  with the 
insulating regime  is the fact that $P_c(\log g_{\perp})$ 
remains system-size invariant.
The relation of Eq. (\ref{var}) is also
valid for the critical points B and D where in
the $\perp$ direction $\langle g_{\perp} \rangle << 1$.

\section {discussion}

The random topological multilayered structure studied
may be regarded as a first step towards an explanation,
via non-interacting electrons,
for properties of strongly anisotropic materials.
For example, in the case of $W=0$ the $2D$ layers
are perfect and the disorder represented by $p$ can
be due to impurities or oxygen vacancies
in the insulating layer among the $2D$ planes of the
cuprates. The electrons propagating in the perpendicular
direction of this system shall encounter anisotropy in the disorder
due to the random interplane links in addition to the value
of $t$ which can be different to that of the parallel
direction.  However,
the  distribution of the critical conductance $P_{c}(g)$
depends both on the choice of the parameters and the
direction where the electron moves.  

\medskip

The considered anisotropic structure
exhibits rather strange transport properties on a given scale,
expressed in the dramatic differences of the critical
conductance in the parallel and perpendicular directions. 
In the perpendicular direction the conductance distribution slightly  
bellow the critical point is log-normal resembling
the statistical properties typical of an insulator.
A similar statistical ``anomaly" has been described in
\cite{20}.  However, a strong difference to a ``true"  insulating regime
exists since the conductance still grows with the size.
In the large size limit the corresponding distribution reaches
a  Gaussian. An analogous discussion holds 
for the parallel direction slightly above the critical point.
The main criterion for the specification 
of the critical regime is the size dependence
of the conductance $\langle g\rangle$ (or $\langle\log
g\rangle$).

\medskip

The proposed model for $W=0$ may have some relation 
to the strongly anisotropic transport proprties
of high--$T_c$ cuprates. 
In these materials, as temperature increases, the
inelastic scattering due to phonons, spin waves or
other excitations within the $CuO$ planes, can cause 
a decrease of the  inelastic scattering length $l_{in}$.
If the temperature is so high that $l_{in}$ becomes smaller
than mean free path the transverse conductivity is metallic.
Experiments for  Bi$_2$Sr$_2$CaCu$_2$O$_8$
and underdoped La$_{2-x}$Sr$_x$CuO$_4$, YBa$_2$Cu$_3$O$_{6+x}$
give  out--of--plane resistivity which has a  semiconductor--like
temperature dependence at low temperatures (high at small--$T$
with a rapid decrease by increasing $T$)  and a linear--in--$T$
behavior at high temperatures. The characteristic
crossover temperature between the two regimes 
$T^{*}$ decreases by increasing the doping in La$_{2-x}$Sr$_x$CuO$_4$
and YBa$_2$Cu$_3$O$_{6+x}$ \cite{7,8,9,10}.
We notice that if we relate the
bond density $p$ with the doping density
of the high-$T_c$ materials  the obtained $p$-dependence of the
perpendicular mean free path can be used to
explain qualitatively the  reported behavior.
It must be pointed out
that the relation between the cuprate doping density  and
the bond density $p$ is natural, since an increase in the
number of the doping impurities
or the oxygen atoms in the layer between two CuO$_2$ planes
increases the number of hopping paths between the two planes.
As $T^{*}$ decreases further (below $T_c$) the out--of--plane
normal--state resistivity also becomes metallic,
which has been observed in high--quality
single crystals of YBA$_2$Cu$_3$O$_7$ and other
high--$T_c$ cuprates corresponding to the absence of
disorder ($p\approx 1$ with $W=0$  in the proposed model) with almost
infinite perpendicular mean free path. This effect occurs
only in the perpendicular direction since the parallel
mean free path is always much longer and inelasting scattering 
becomes dominant.

\medskip

In summary, we have introduced a simple layered lattice model
with anisotropic disorder  described by
the interplane bond density $p$, in addition to the usual
anisotropic band structure expressed
via the interplane hopping $t$.
In the absence of diagonal disorder we show  extended states in both
directions but the obtained
mean free path and the conductance in the ${\perp}$
direction is much smaller than in the
${\parallel}$ direction. Moreover, $g_{\perp}$
fluctuates strongly  as a function of
energy, which leads to an insulator-like
temperature dependence of the conductivity in the $\perp$ direction.
In the presence of additional diagonal disorder
of strength $W$ we have shown that the critical disorder
and the critical exponent $\nu$ do not depend 
on the transport direction. 
The obtained data for the localization exponent $\nu$
agree with recent accurate estimates for the isotropic model
$\cite{23}$ and confirm the universality at the metal-insulator transition.
The obtained critical conductance distribution $P_c(g)$ 
although independent on the system size
depends strongly on the parameters 
and the direction of transport.
\medskip

{\bf Acknowledgments}

This work was supported in part  by a TMR network.
SNE and SJX thank a Chino-Greek grant  and
PM the Slovak Grant Agency and  NATO.  
We also like to thank Professors Xing 
and Economou for many useful discussions.

 {}

\newpage\clearpage
%
%
%
%
\par
\centerline{\psfig{figure=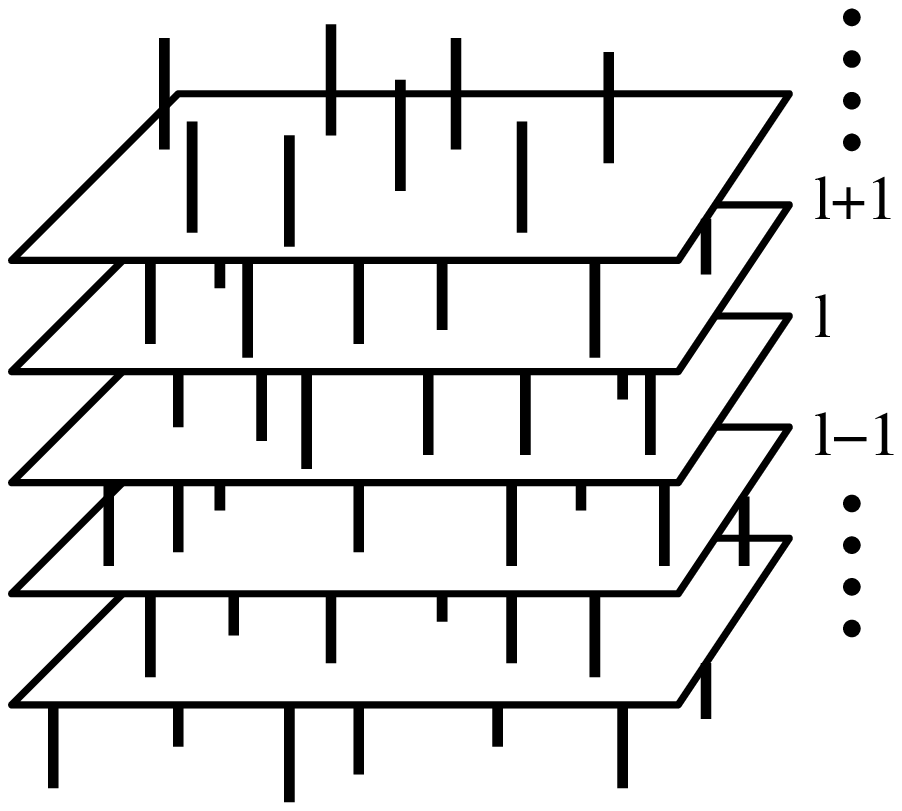,height=2.3in}}
{\footnotesize{{\bf FIG. 1.}
 A picture of the multilayered structure which consists of
 $2D$ square lattices $l$ (layers) connected by perpendicular bonds of
 strength $t$ placed at random positions with probability $p$.
}}
\par
\vspace{.4in}
%
\par
\centerline{\psfig{figure=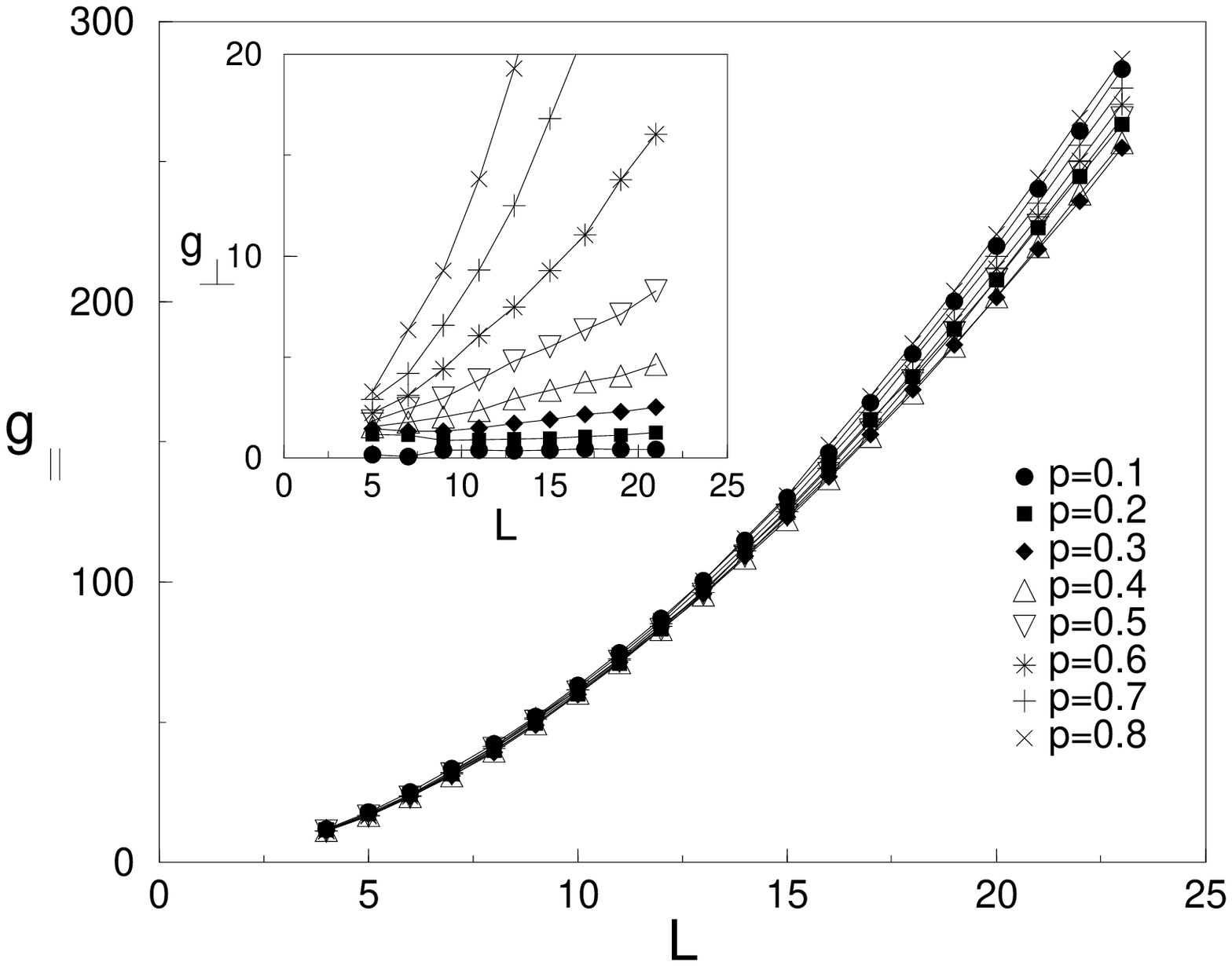,height=2.3in}}
{\footnotesize{{\bf FIG. 2.}
The $g_{\parallel} $ as a function of the linear system size $L$ for a
cubic $L\times L\times L$ system of parallel planes with $W=0$
and randomly placed interplane bonds of
density $p$ with strength $t$=0.3. 
In the inset $g_{\perp}$ for the
same system exhibits similar behavior but much smaller
values.
}}
\par
\vspace{.4in}
%
%
\par
\centerline{\psfig{figure=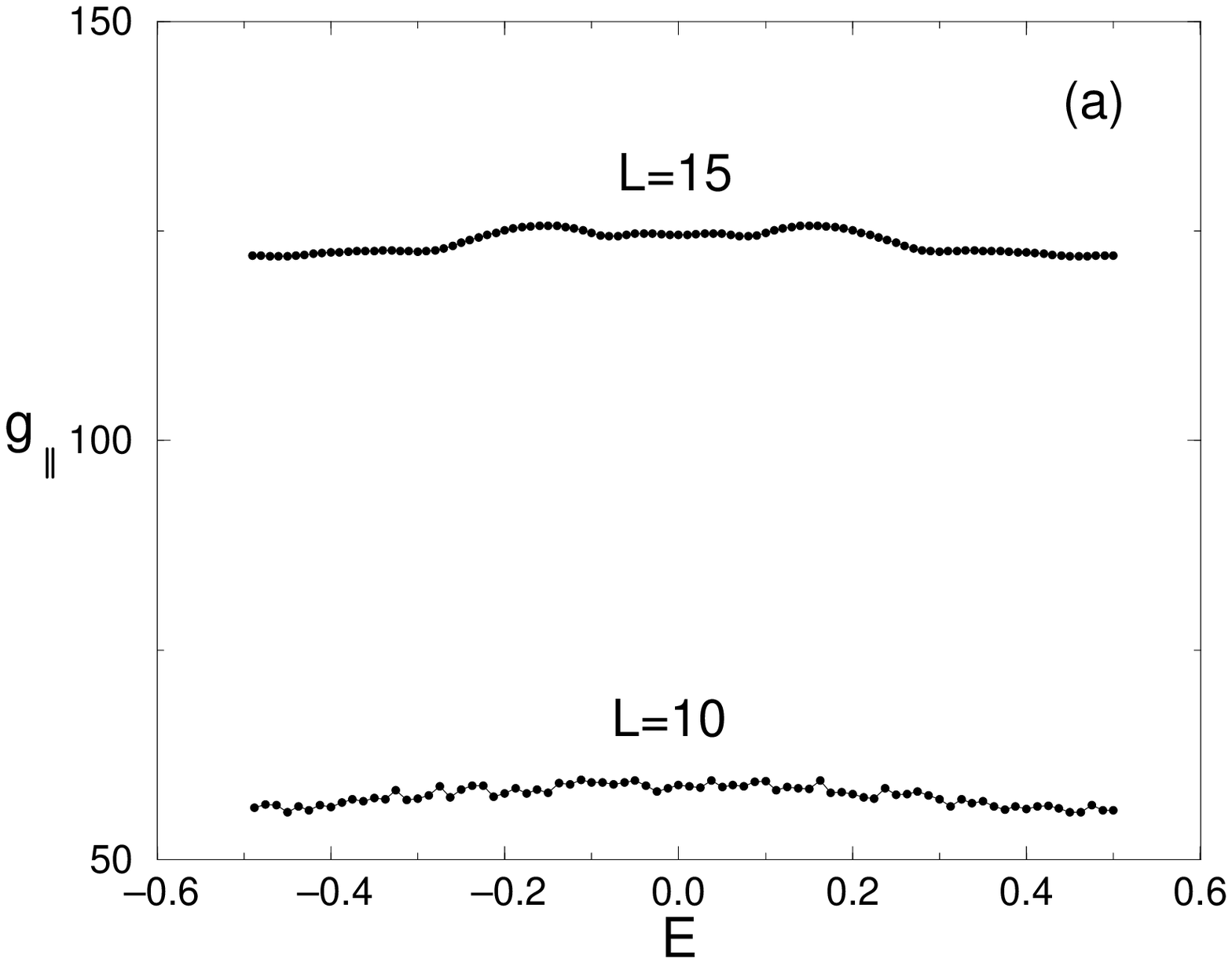,height=2.3in}}
\vspace{.1in}
\centerline{\psfig{figure=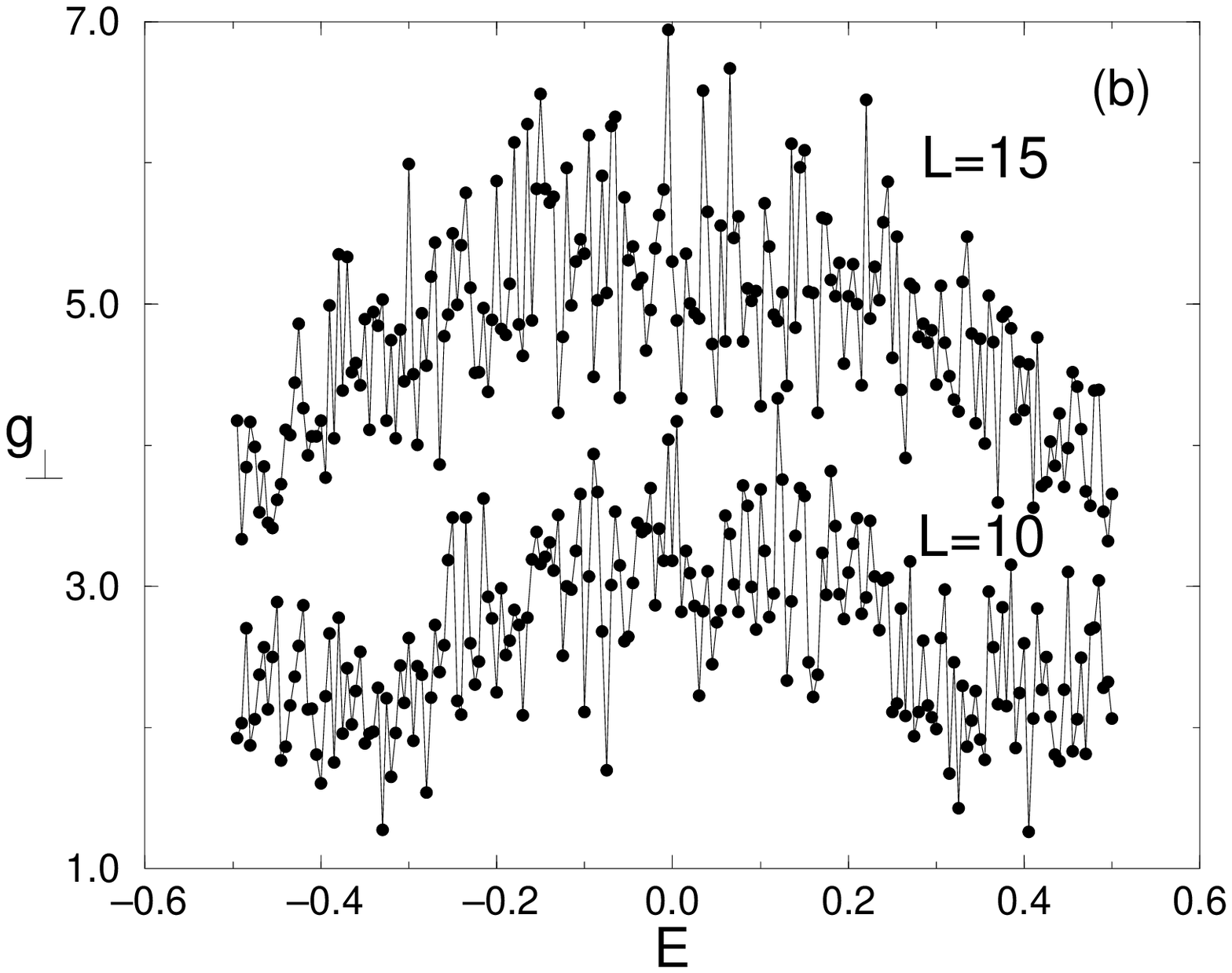,height=2.3in}}
{\footnotesize{{\bf FIG. 3.}
{\bf (a)} The energy-dependent $g_{\parallel}$ for a cubic layered
system with $L$=10,15 and $W=0$, $t$=0.3, $p$=0.5. {\bf (b)} The 
$g_{\perp}$ is much smaller and exhibits violent oscillations
as a function of the energy $E$.}}
\par
\vspace{.4in}
%
%
\par
\centerline{\psfig{figure=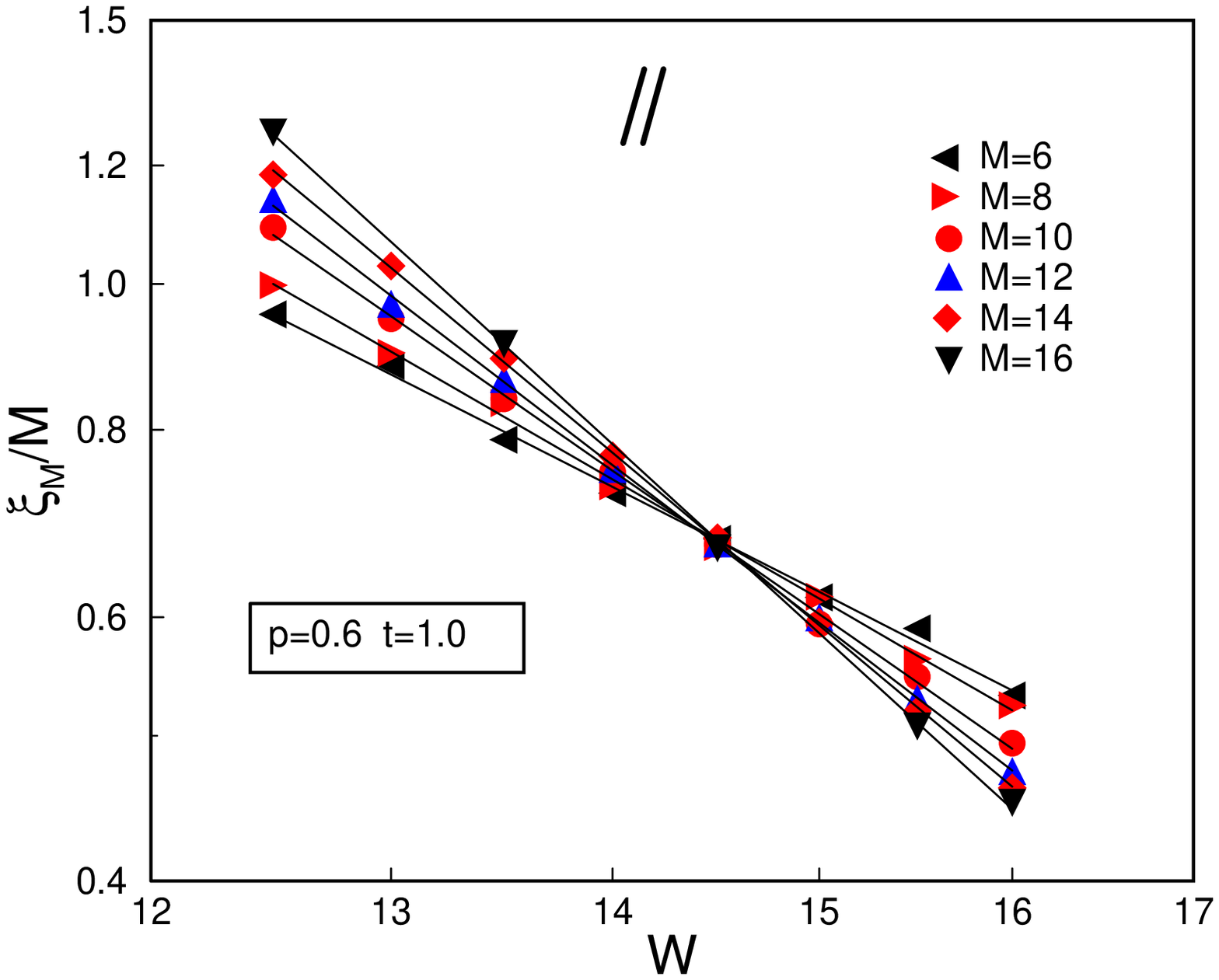,height=2.3in}}
\vspace{.1in}
\par
\centerline{\psfig{figure=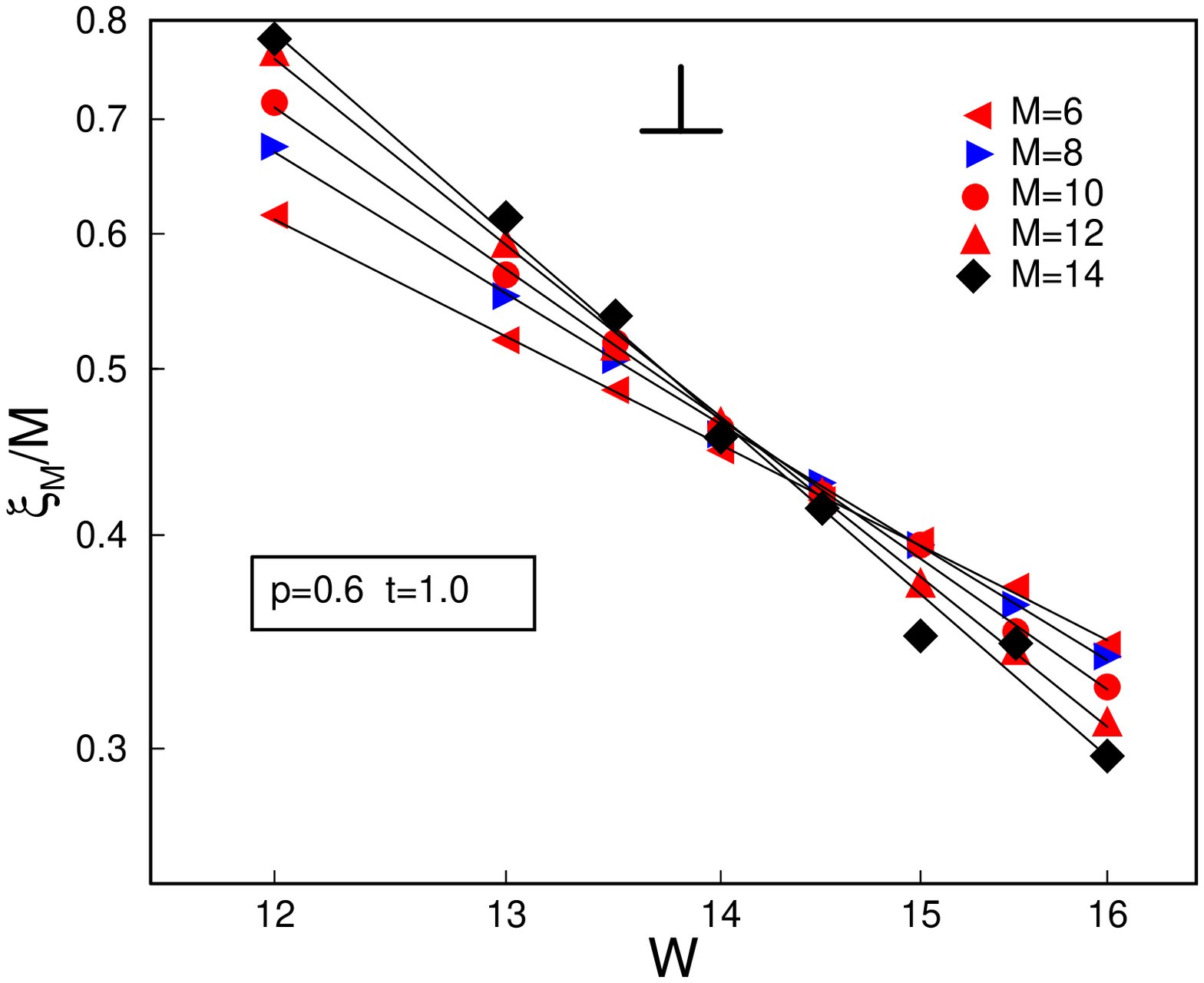,height=2.3in}}
{\footnotesize{{\bf FIG. 4.} {\bf {A:}}
The behavior of the scaled localization length $\xi_M/M$ 
for the parallel and perpendicular direction
in the $M\times M\times L$ system
with $p=0.6$ and  $t=0.1$ as a function of $W$.  
The critical point is located at $W_{c}\approx 14.47$ 
in the ${\parallel}$ direction
and $W_{c}\approx 14.30$ in the
${\perp}$ direction.
}}
\par
\vspace{.3in}
%
\par
\centerline{\psfig{figure=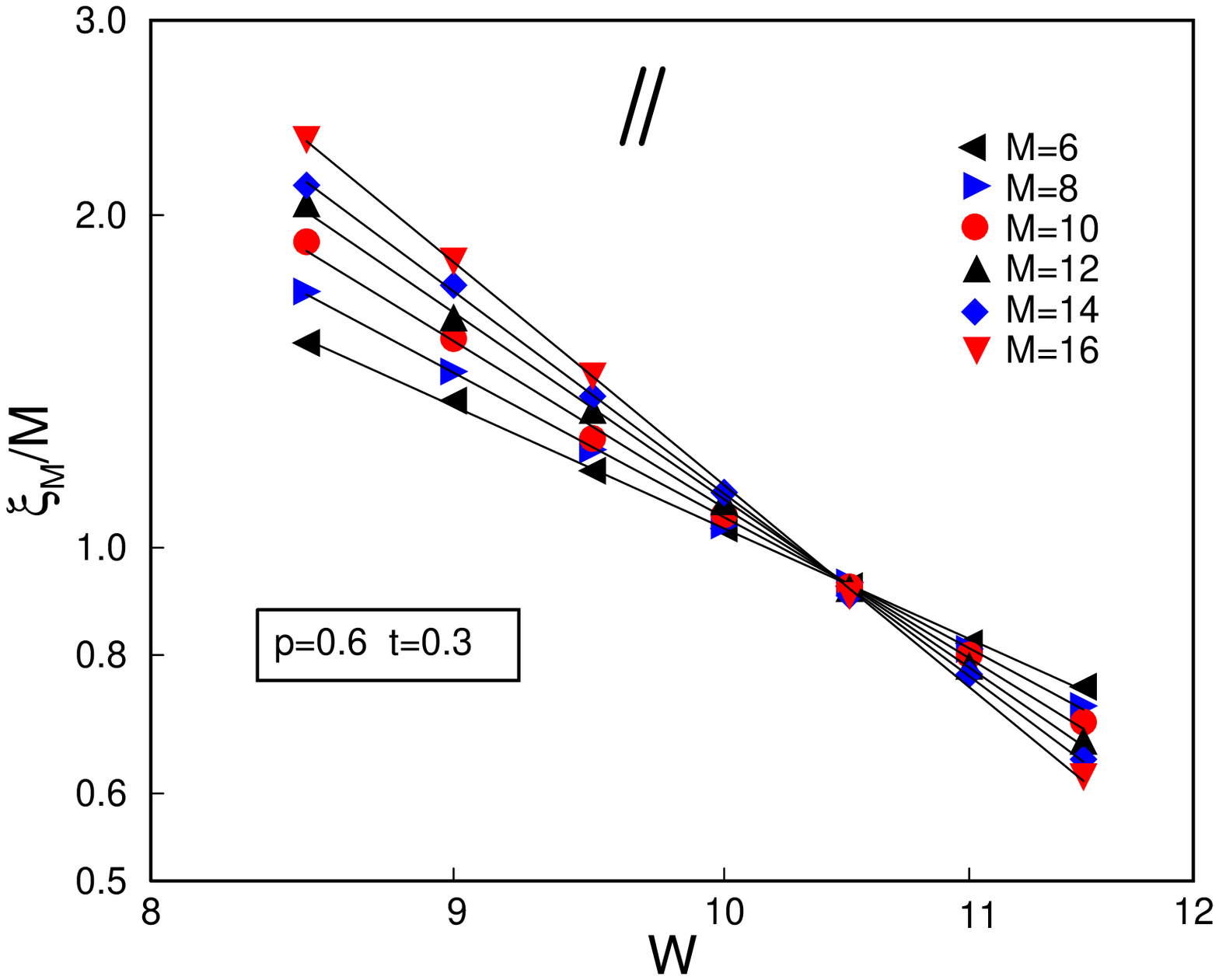,height=2.3in}}
\vspace{.1in}
\centerline{\psfig{figure=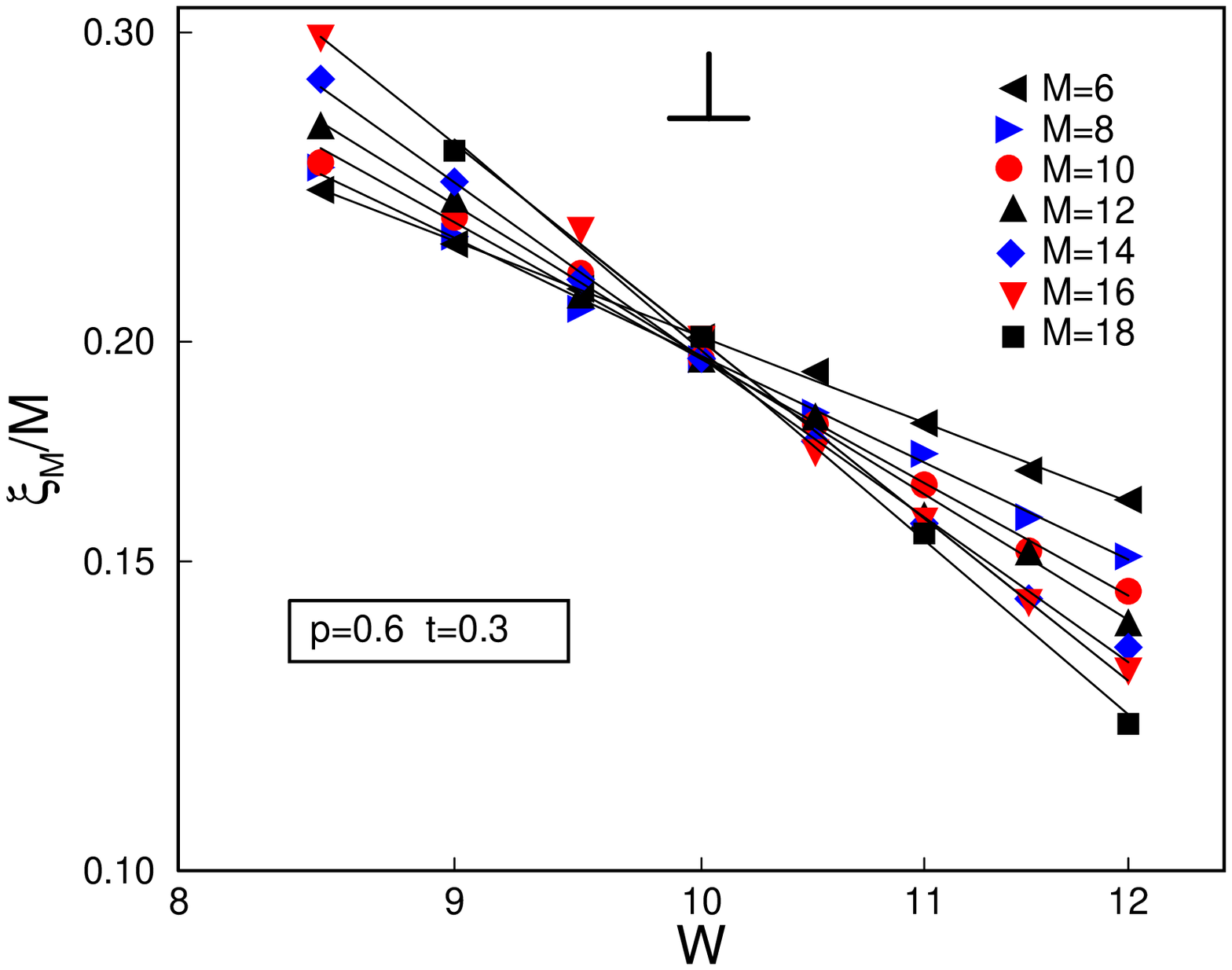,height=2.3in}}
\vspace{.1in}
{\footnotesize{{\bf FIG. 5.} {\bf {B:}}
The behavior of the scaled localization
length $\xi_M/M$ for the parallel and perpendicular direction
in the $M\times M\times L$ system
with $p=0.6$ and  $t=0.3$ as a function of $W$.  The critical point
is located  at $W_{c}\approx 10.48$ in the ${\parallel}$ 
and $W_{c}\approx 10.20$ in the ${\perp}$ direction.
}}
\par
\vspace{.3in}
%
%
\par
\centerline{\psfig{figure=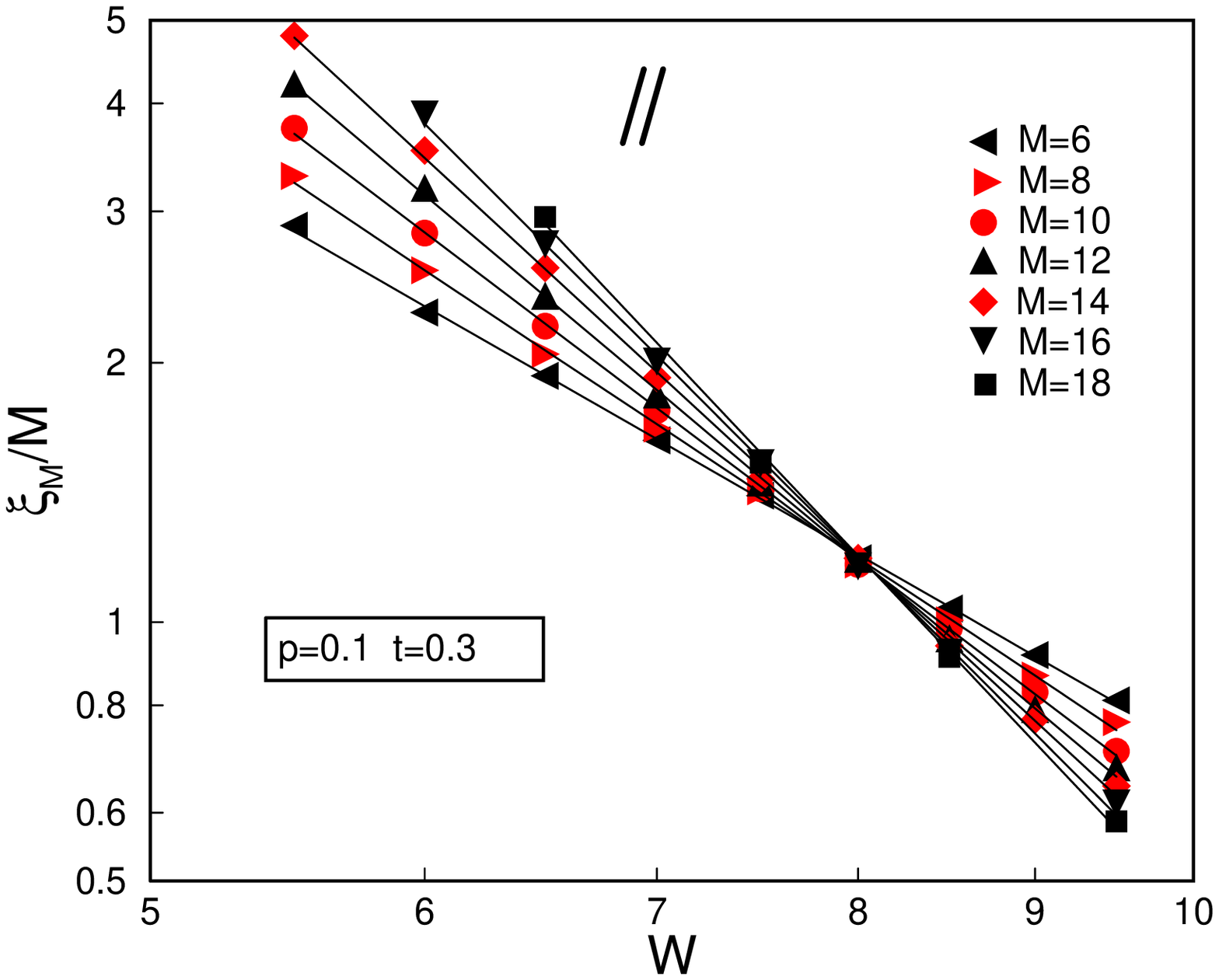,height=2.3in}}
\vspace{.1in}
\centerline{\psfig{figure=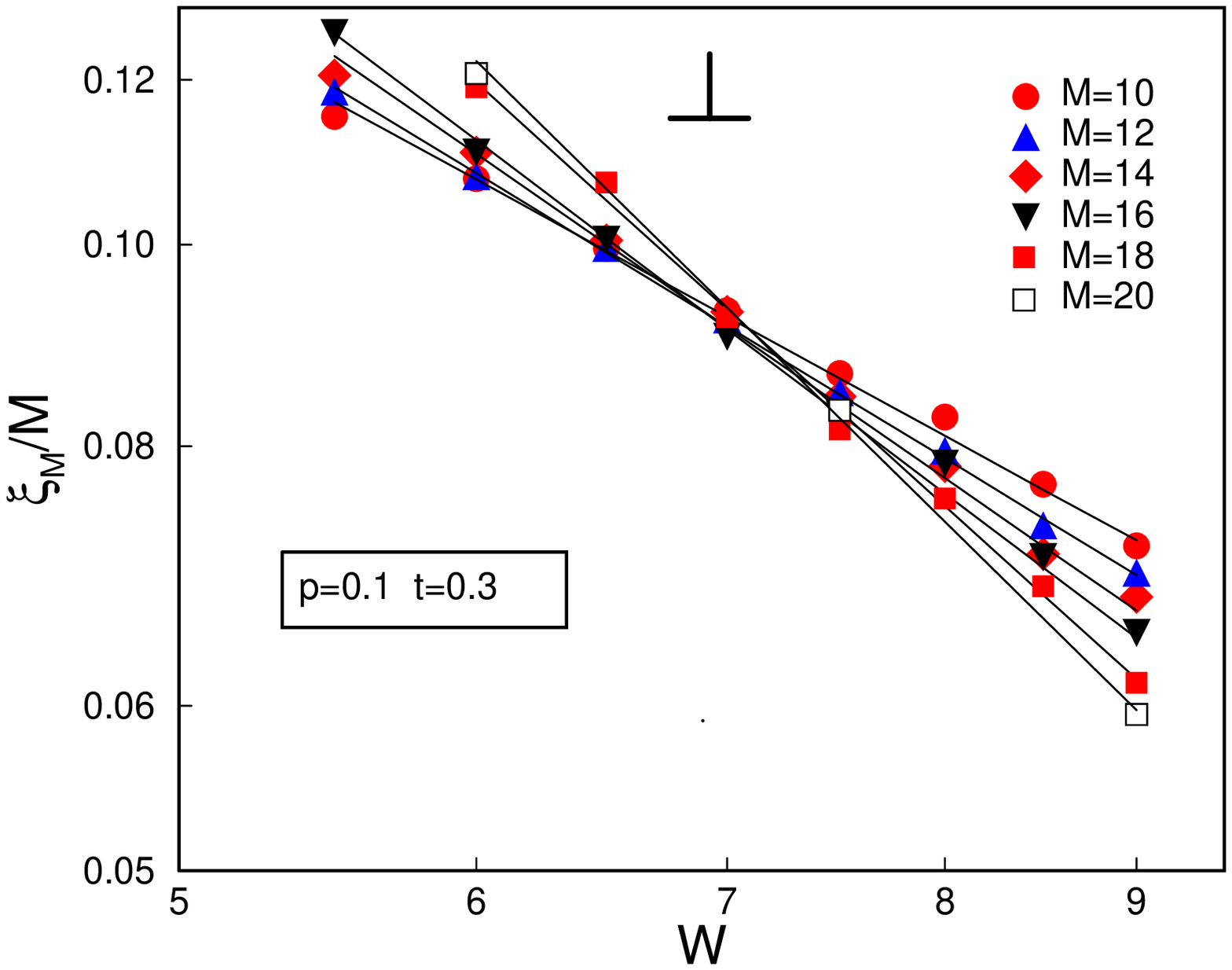,height=2.3in}}
\vspace{.1in}
{\footnotesize{{\bf FIG. 6.} {\bf{C:}}
The behavior of the scaled localization
length $\xi_M/M$ for the parallel and perpendicular direction
in the $M\times M\times L$ system
for $p=0.1$ and  $t=0.3$ as a function of $W$.  
The critical point is displayed at $W_{c}\approx 7.93$ 
in the ${\parallel}$ and $W_{c}\approx 7.18$
in the ${\perp}$ direction.
}}
\par
\vspace{.3in}
%
%
%
\par
\centerline{\psfig{figure=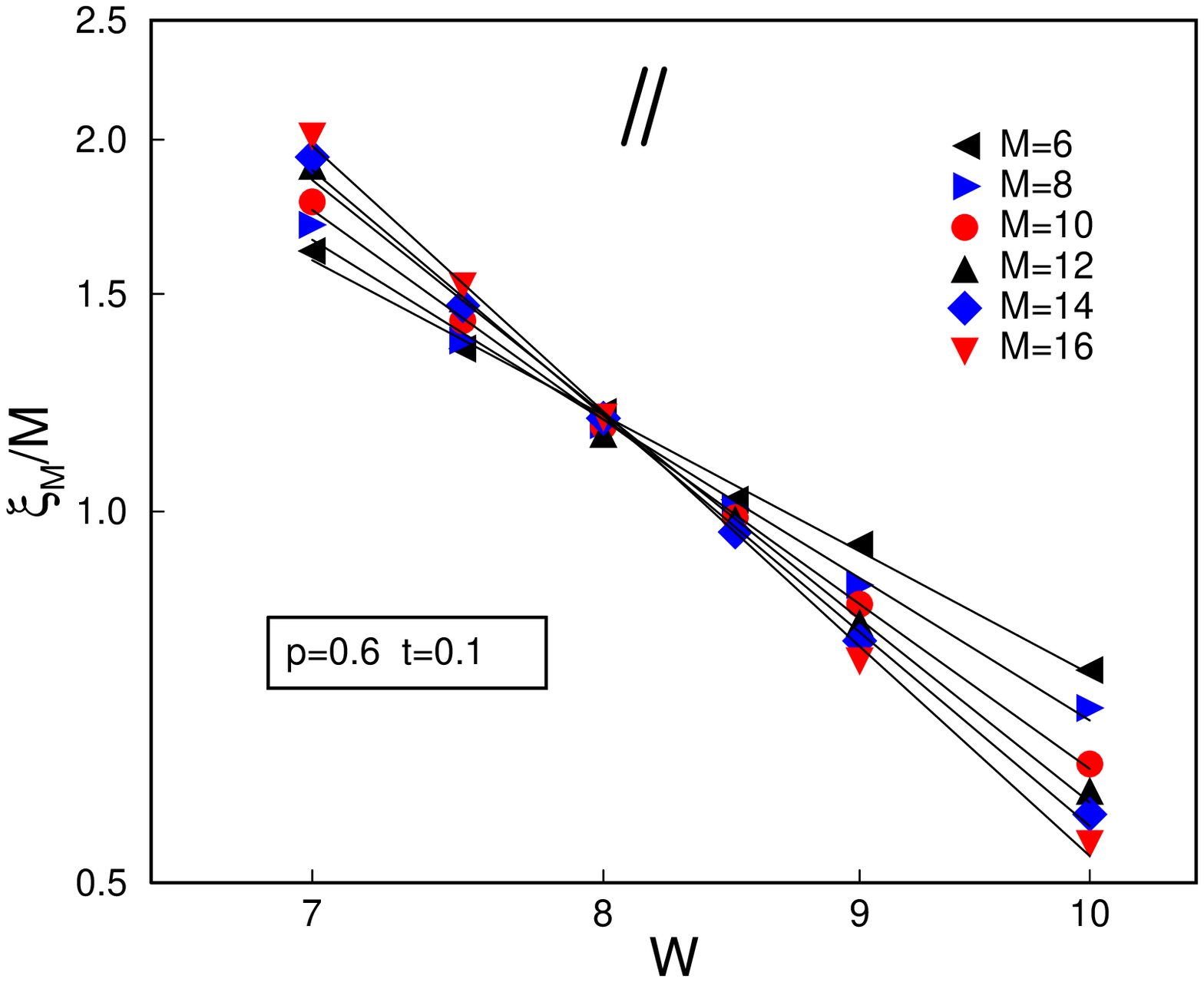,height=2.3in}}
\vspace{.1in}
\centerline{\psfig{figure=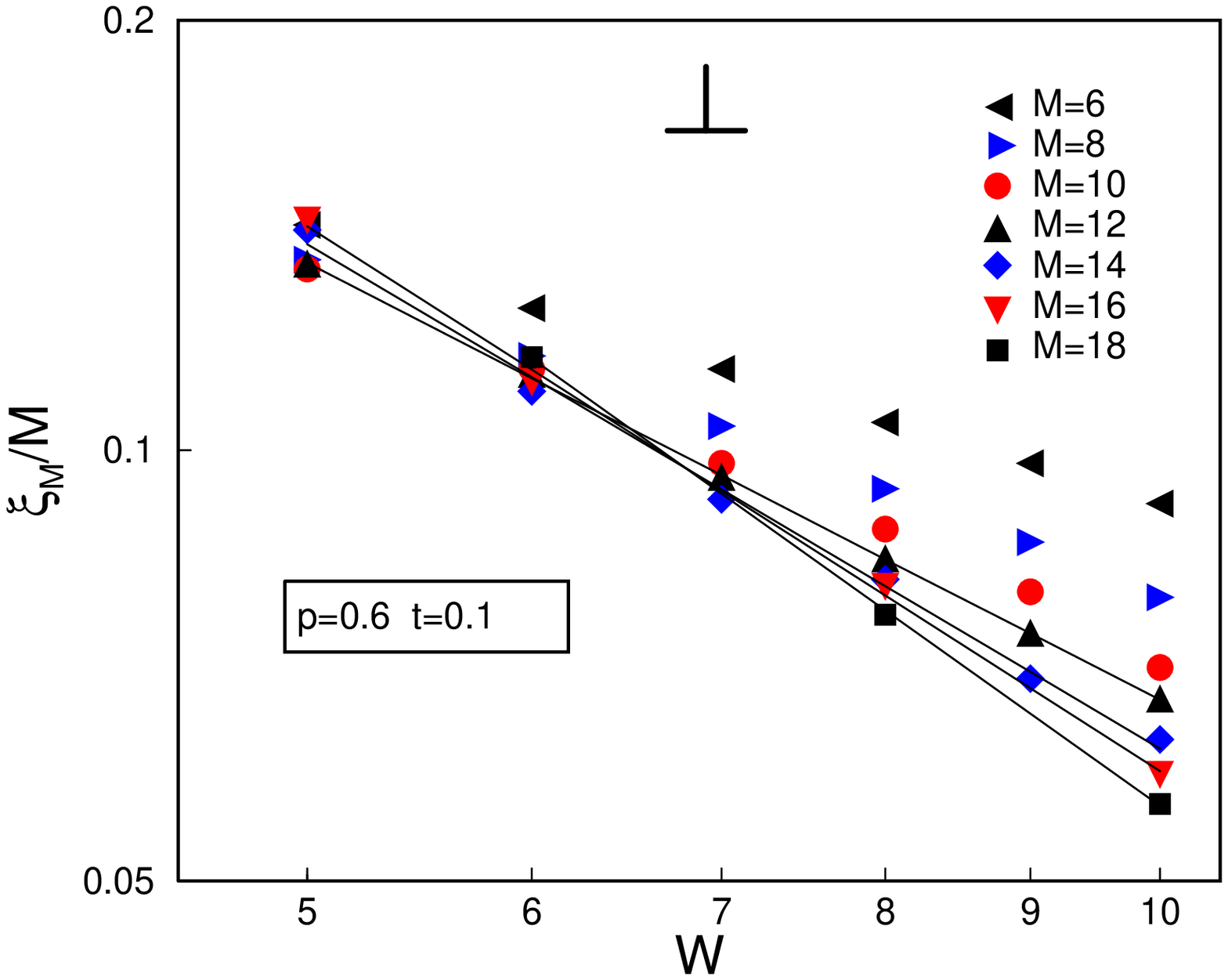,height=2.3in}}
\vspace{.1in}
{\footnotesize{{\bf FIG. 7.} {\bf {D:}}
The behavior of the scaled localization length
$\xi_M/M$ for the ${\parallel}$ 
and ${\perp}$ direction in the $M\times M\times L$ system
with $p=0.6$ and  $t=0.1$ as a function of $W$.
The critical point is located in $W_{c}\approx 8.05$ 
in  the  ${\parallel}$ and $W_{c}\approx 6.80$
in  the ${\perp}$ direction.
It is seen that the data for smaller $M=6,8$ 
fail to cross at the same point indicating ``insulating" 
behavior.}}
\par
\vspace{.3in}
%
\par
\centerline{\psfig{figure=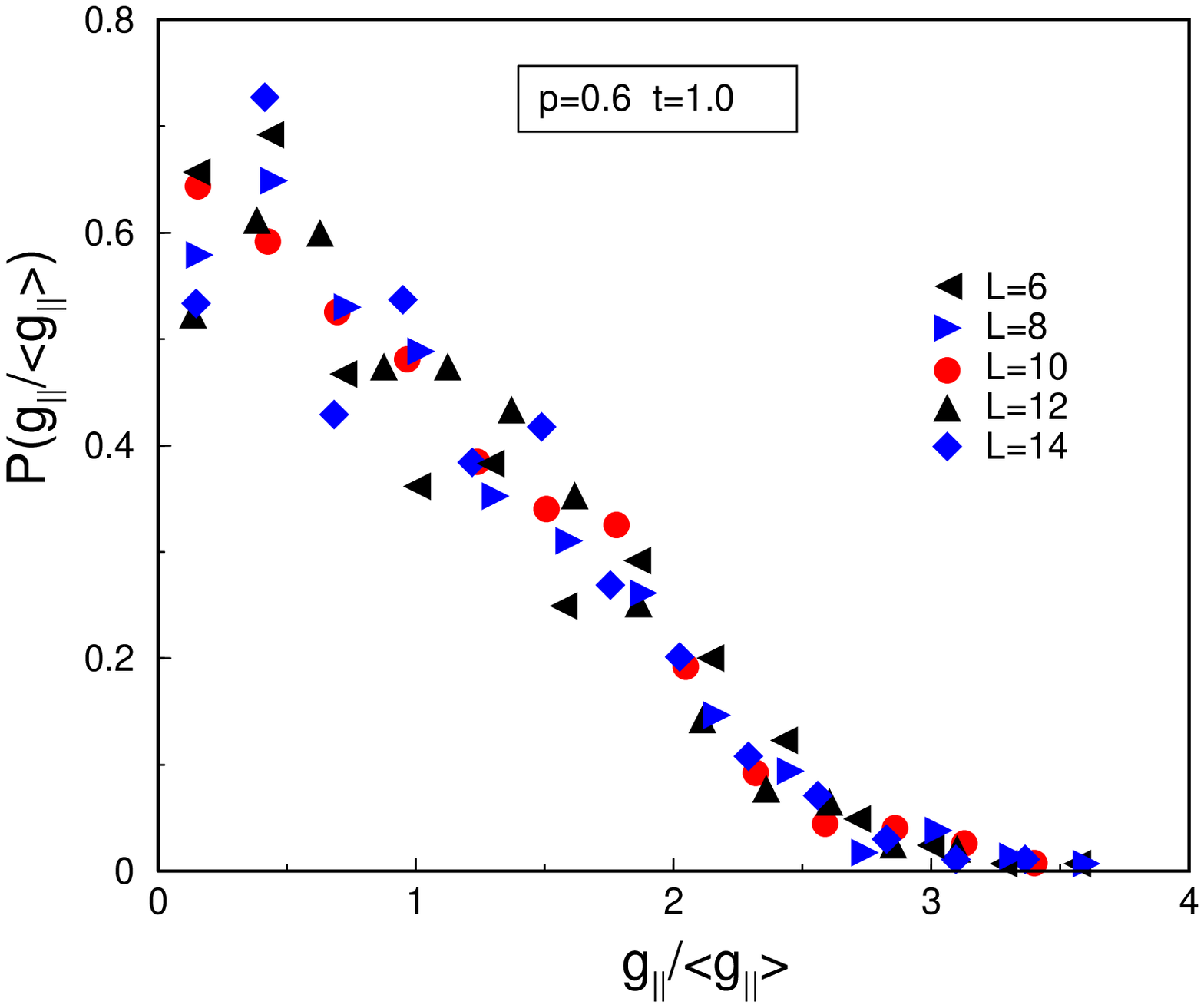,height=2.3in}}
\vspace{.1in}
\centerline{\psfig{figure=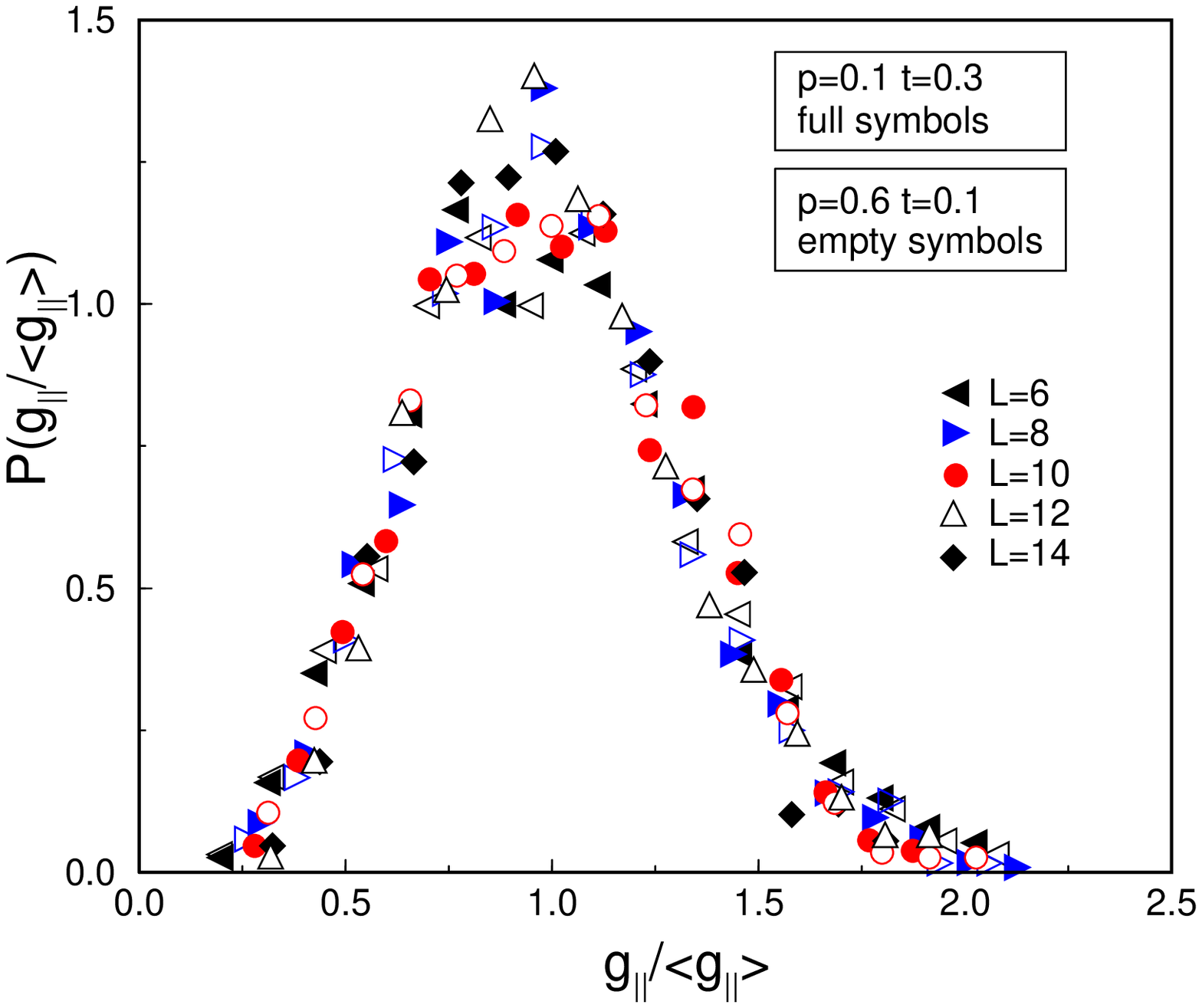,height=2.3in}}
\vspace{.1in}
\centerline{\psfig{figure=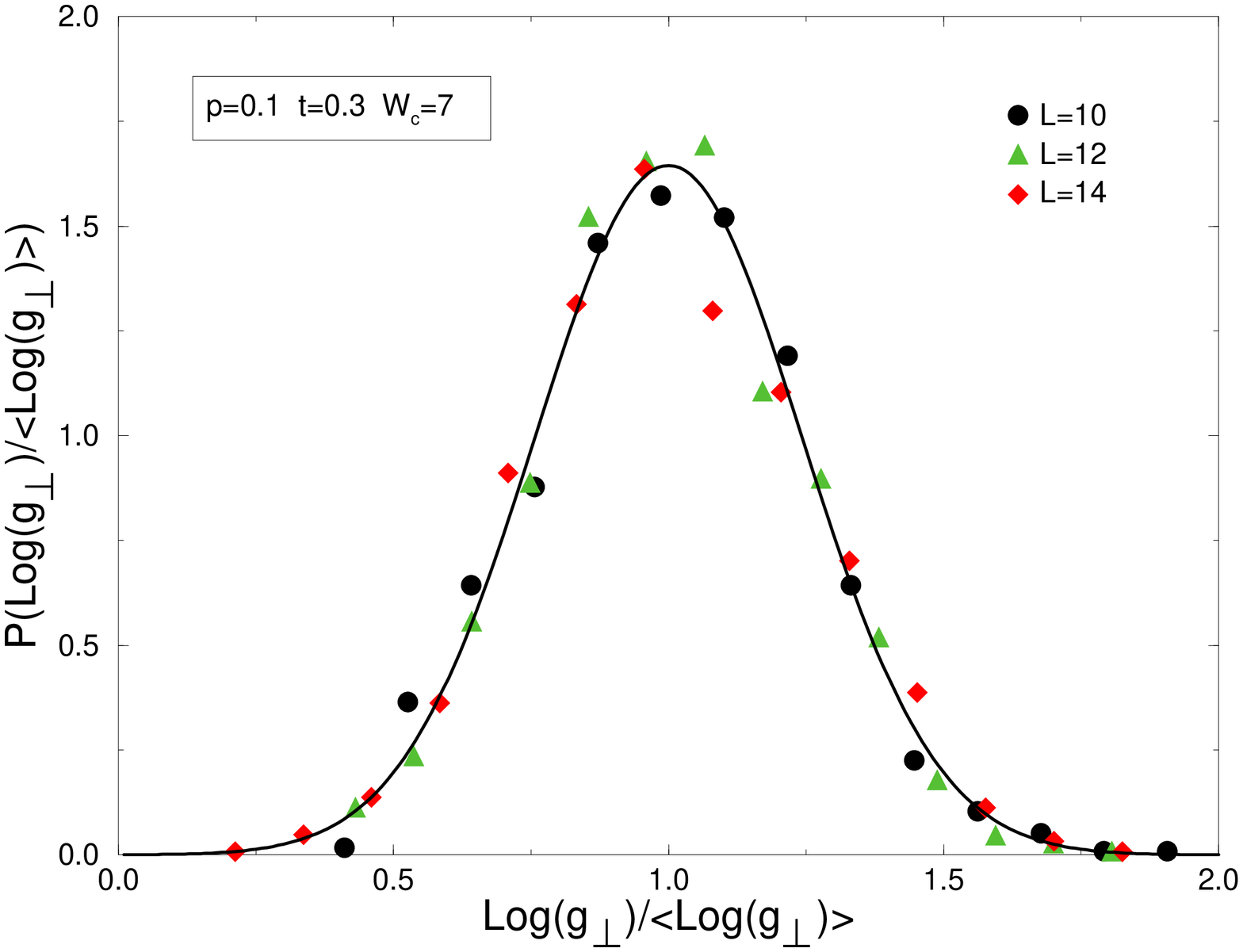,height=2.08in,angle=-90}}
{\footnotesize{{\bf FIG. 8.}
{${\bf (a)}$. The critical distribution of the parallel conductance 
in the case A with $<g_\parallel>\approx 0.59$, var$ (g_\parallel)=0.16$.
${\bf (b)}$.  The same as in (a) for the critical points
C (full symbols) and D(open symbols) with $<g_\parallel>\approx 2.6$,
var $(g_\parallel)=0.7$. The C,D have the same $\Lambda_c$
and the same critical distribution (see Table 1).
${\bf (c)}$.  The  critical distribution
of $P(\log g_\perp)$ in the perpendicular direction 
for case C is also shown for comparison.}
}}
\end{document}